\documentclass[conference]{IEEEtran}
\IEEEoverridecommandlockouts
\usepackage{cite}
\usepackage{amsmath,amssymb,amsfonts}
\usepackage{algorithmic}
\usepackage{graphicx}
\usepackage{textcomp}
\usepackage{xcolor}
\usepackage[english,german]{babel}
\def\BibTeX{{\rm B\kern-.05em{\sc i\kern-.025em b}\kern-.08em
    T\kern-.1667em\lower.7ex\hbox{E}\kern-.125emX}}
\begin{document}
\selectlanguage{english}	
\title{Hyper Space Exploration \linebreak A Multicriterial Quantitative Trade-Off Analysis for System Design in Complex Environment\\
}

\author{\IEEEauthorblockN{1\textsuperscript{st} Herbert Palm}
\IEEEauthorblockA{\textit{University of Applied Sciences Munich} \\
\textit{Systems Engineering}\\
Munich, Germany
\\
herbert.palm@hm.edu}

}

\maketitle

\begin{abstract}
Successful engineering requires environmentally adapted procedural and architectural approaches. While dealing with complicated issues has become an engineering standard mastering uncertainties in complex environment is still a major issue. Global trends, such as an increasing rate of disruptive (non-evolutionary) technology changes or merging of technology fields, however, enforce the importance of this complex habitat. \newline Missing experience in a priori unknown technological territory faces engineers with two questions of paramount importance: \newline 1) How can the best (rather than a first) architectural solution within the space of potential alternatives be identified? \newline 2) How can a proof-of-concept for any considered solution prior its implementation be provided?  Mastering lack of knowledge related risks and uncertainties states one of the most prominent tasks in according projects. The paper presents a novel approach of a system design methodology in such complex environment called Hyper Space Exploration (HSE). \newline The HSE approach combines methods of virtual prototyping with those of design of virtual experiments based studies for statistical learning. Virtual prototyping allows an early feedback on system behavior with a proof-of-concept prior implementation. Statistical learning enables system architects to systematically build up the space of potential solution alternatives, model the effects of design and use-case variables on target indicators in complex territory, quantify target indicator trade-offs and finally identify Pareto-optimal system solutions.  \newline The first part of the paper characterizes engineering challenges in complex environment. Section two presents the HSE methodology with its two major constituents work flow (part A) and tool chain (part B). Section three outlines first successful HSE applications that already proved its capabilities and universality. Final section four gives an outlook to further HSE applications as well as methodological future HSE extensions.
\end{abstract}


\section{System Design in Complex Environment}
\begin{figure}[htbp]
\centerline{\includegraphics[width=7cm]{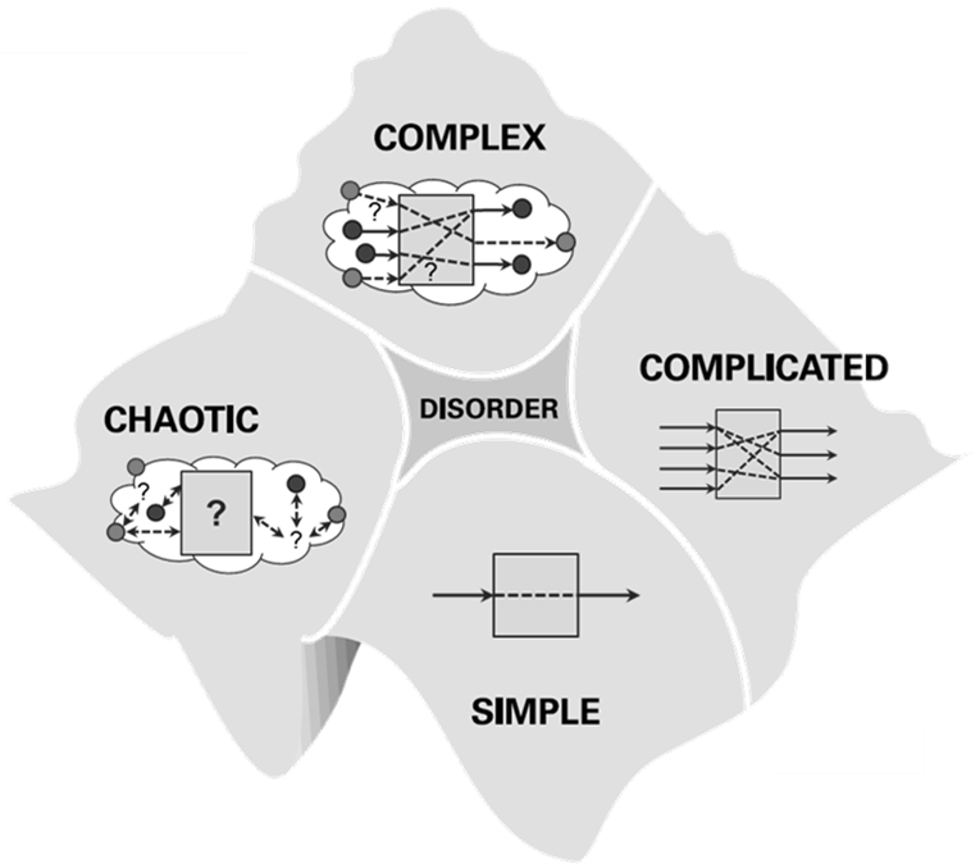}}
\caption{Transfer of the \textit{Cynefin framework} \cite{Snow2007} to engineering.}
\label{fig_cynefin_for_engineers}
\end{figure}

In 2007 David Snowden and Mary Boone presented ``a leader's framework for decision making'' \cite{Snow2007}. The proposed framework categorizes decision making processes according to a problem's cynefin (Welsh for habitat or environment). Knowing the  ``prevailing operative context" of a given problem enables leaders to choose the right sequence of actions: While \textit{simple} or \textit{complicated} problems  ``where cause-and-effect relationships are perceptible'' favor a categorization or analysis focus to identify best or good practice solutions \textit{complicated} or \textit{chaotic} problems where ``there is no immediately apparent relationship between cause and effect, and the way forward is determined based on emerging patterns'' focus has to be put on (feedback based) probing or on acting itself. Snowden and Boone's \textit{Cynefin framework} has been primarily addressed to general management rather than to the engineering community. It has been  adopted, however, meanwhile to a systems engineering adequate and consistent terminology (see Fig. \ref{fig_cynefin_for_engineers}).

Translating the original \textit{Cynefin framework} requests engineers to assign given problems to one of the following habitats:
\begin{itemize}
\item \textit{simple} problems are characterized by few linear cause-effect relations causing  a known dynamic behavior. Lean procedural approaches such as PDCA \cite{Dem1986} are well suited to solve \textit{simple} problems on base of known best practice.  
\item \textit{complicated} problems are characterized by non-linear or a large number of cause-effect relations. Predictable behavior is achievable but requests thorough analysis. Subsequent target setting, design, implementation and test procedural approaches such as waterfall or learning cycle \cite{Boehm1979} procedural models are well suited to solve \textit{complicated} problems.  Issues caused by a large number of to be reflected cause-effect relations may, in addition, be managed by the hierarchic approach of the V-Model \cite{Fors1998}. Introducing hierarchy layers tames complicacy caused by multiplicity of cause-effect relations. 
\item \textit{complex} problems are characterized by a dynamically changing topology (e.g. caused by open system borders) or not fully \textit{a priori} predictable cause-effect relations. Lack of knowledge when designing solutions for \textit{complex} problems usually manifests in two ways: a) Missing technical experience (or unclear requirements) request a proof-of-concept (system feedback) and b) inability of instantaneously picking the most effective and efficient solution requests a trade-off comparison amongst potential solutions. For both issues procedural approaches have been identified: Virtual prototyping \cite{Schaaf1997} approaches provide early system feedback and may serve as a proof-of-concept based validation. On the other hand, macro procedural approaches such as hierarchic studies \cite{Daenzer2002} or tradespace explorations \cite{Ross2005} are well suited to systematically build up the space of potential solutions in order to subsequently search for most favorite solutions.  
\item \textit{chaotic} problems are characterized by a predominant lack of knowledge on system topology (e.g. by rapidly varying constituents or their interconnects) or cause-effect relations. Confinement of a  \textit{chaotic} system to subsystems of remaining \textit{complex}, \textit{complicated} or \textit{simple} character currently seems the only prevailing approach suited to solve \textit{chaotic} problems. 
\end{itemize}
Choosing a habitat adequate sequence of actions to solve an engineering problem bears a fundamental benefit: It enables solving a problem effectively while simultaneously using required resources efficiently. The \textit{Cynefin for engineers framework} allows to decide on useful front-loading measures (i.e. enforcing activities in early life-cycle phases). Front-loading  does not bear any advantage in itself. It rather ponders the value of managing uncertainties and risks against its time and cost related efforts.
\newline
Digitalization or Industry 4.0 represent global trends \cite{Dobbs2015} of disruptive (non-evolutionary) technology changes or merging of technology fields. Related projects will confront engineers more than ever with open systems (e.g. dynamically changing production resources) or not fully predictable cause-effect relations. The HSE approach focuses on this \textit{complex habitat} enabling system architects to master related challenges.

\section{HSE Methodology}
Modeling and simulation form adequate means to analyze capabilities of a system design versus system requirements. The Systems Engineering (SE) Handbook \cite{SE2015}, refines the analysis purpose when stating ``analysis is more than simply determining if the criteria are met but also the degree to which they are met (or fall short or exceed), as this information is used to support trade-offs and evaluation of alternatives". 
Various terms have been established meanwhile with respect to the requested systematic analysis of solution alternatives: Authors of the NASA Systems Engineering Handbook \cite{NASA2007} denote them as ``trade studies" or ``trade-off studies" indicating usual target conflicts (trade-offs) to be reflected. Ross \cite{Ross2005} uses the term ``tradespace exploration" for the analysis of design alternatives in aerospace applications. Design alternatives are ususally characterized by means of a morphological analysis \cite{Zwicky1969}. The \textit{design space} (synonym to \textit{tradespace}), i.e. the space of potential design alternatives, is spanned by all to be considered morphological system layouts. Methodological approaches (\textit{Design Space Exploration} or synonymic \textit{Design Space Evaluation}, DSE) for systematic analysis of design spaces \cite{Kang2010} or \textit{real options} \cite{Neufville2010} for decision making amongst design alternatives are widely spread in the aerospace industry. Surprisingly, established trade-off analysis approaches do not make systematic use of virtual prototypes. In contrast, expert interviews assign a \textit{utility} as well as a \textit{life cycle cost} value to individual design layouts. Trade-offs are then phrased in terms of these selected target indicators utility vs. life cycle cost. \textit{Pareto-optimal}  design solutions (i.e. solutions that may only be improved with respect to one target indicator  when accepting deterioration with respect another target indicator) may be identified. Multicriterial or use-case specific individual trade-offs are usually neglected.
 
\subsection{Generic HSE Work Flow}
\begin{figure}[htbp]
\centerline{\includegraphics[width=7cm]{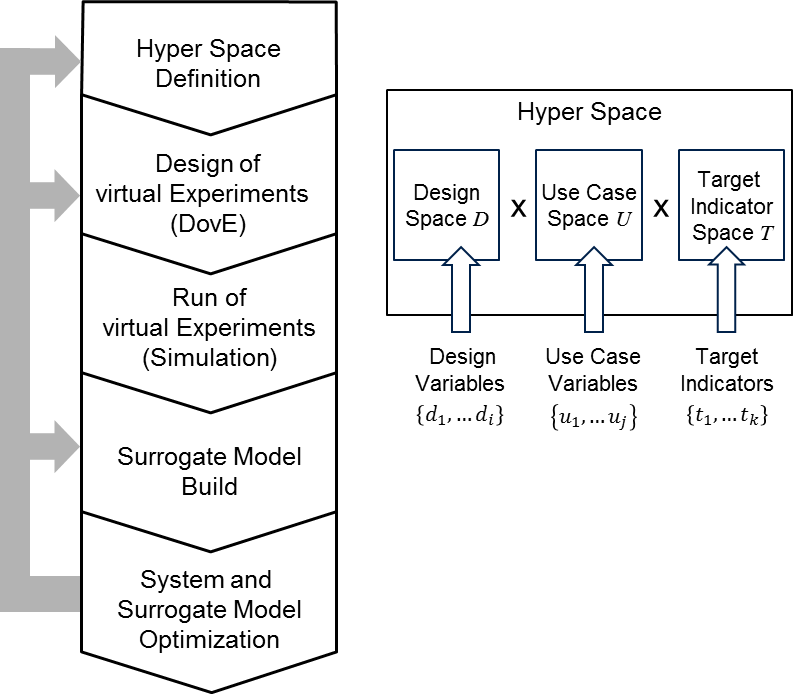}}
\caption{Generic HSE work flow sequence (left), subspaces of the Hyper Space (top right) and basic functionality of the Surrogate Model (lower right).}
\label{fig_generic_HSE_work_flow}
\end{figure}

 HSE extends the DSE approach in several significant aspects based on a 5-step generic work flow as indicated in Fig.~\ref{fig_generic_HSE_work_flow}

Each one of the five steps addresses specific aspects:
\begin{enumerate}

\item \textit{Hyper Space Definition}: Design solutions are characterized by a set of relevant topological and parametric design variables $ \{ d_1, d_2,..., d_i \}$ spanning the i-dimensional \textit{Design Space} $D$. Any vector $d = \left( d_1, d_2,..., d_i \right) \in D$ is representing an individual design layout. Any use use of relevance is characterized by a set of use case variables $ \{ u_1, u_2,..., u_j \}$ spanning the j-dimensional \textit{Use Case Space} $U$. Any vector $u = \left( u_1, u_2,..., u_j \right) \in U$ is representing an individual use case.  Any relevant target indicator for evaluation of design alternatives is characterized by a set of target indicator variables $ \{ t_1, t_2,..., t_k \}$ spanning the k-dimensional \textit{Target Indicator Space} $T$. Any vector $t = \left( t_1, t_2,..., t_k \right) \in T$ is representing an individual multicriterial target achievement measure. The \textit{Hyper Space} $H$ is defined as the Cartesian product of \textit{Design Space}, \textit{Use Case Space} and \textit{Target Indicator Space} $D \times U \times T$.   

\item \textit{Design of virtual Experiments (DovE)}: HSE is exploring unknown technological territory based on a systematically built up set of virtual prototypes suited for subsequent statistical learning. Dating back to the ideas of Ronald Fisher \cite{Fisher1934} space filling algorithms are applied accordingly for choosing effective and efficient candidates for virtual (i.e. simulation based) experiments. From space filling point of view any \textit{Design Space} vector may be treated the same way as any \textit{Use Case} vector. DovE results in an experimental test plan.

\item \textit{Run virtual Experiments (Simulation)}: Virtual experiments, i.e. simulations, are carried out according to the experimental test plan. The run sequence usually will be automated on a script base. Results of simulations are filed in a storage allowing to assign $(d,u)$ based virtual experiments to their according target measures $t$.  

\item \textit{Surrogate Model Build}: System behavior is manifested in the $t(d, u)$ relation. However, this relationship in most cases is analytically inaccessible. Therefore, a surrogate model \cite{Box2005} \cite{Forr2008} in terms of a k-dimensional vector $\hat s$ (of a suited functional family $f_{\alpha}(d,u)$ with parameter vector $\alpha$) is extracted approximating the $t(d, u)$ relation:
\begin{equation}
\hat s= f_{\alpha}(d, u) \approx t(d, u) 
\label{eq:surrogat-formel}
\end{equation}

$\hat s$ represents the surrogate model by assigning a target indicator $t$ approximation to each individual design alternatives $d$ when being applied within a use case $u$. The surrogate model maps the Euclidean product of \textit{Design Space} and \textit{Use Case Space} to the \textit{Target Indicator Space}:
\begin{equation}
D \times U  \overset{\hat s}\to T
\label{eq:hyperspaceabbildungs-formel}
\end{equation}

providing an analytically accessible approximation with quantifiable error within a chosen validation area.

\item \textit{System and Surrogate Model Optimization}: The surrogate model enables system architects now to quantify target indicator trade-offs. It enables assessment of design family capabilities in terms of use case $u$ specific Pareto-optimal solution alternatives:
\begin{equation}
P_u := \{ d \in D \mid \nexists \: d^\prime \in D:  f_{\alpha}(d, u) \prec f_{\alpha}(d^\prime,u)  \} 
\label{eq:pareto-optimal-formel}
\end{equation}

It is also up to the system architect to decide if a reached model accuracy is sufficient to meet requirements for proof-of-concept. If higher accuracy is required the model may be refined a)  by newly defining the chosen Hyper Space segment (i.e. loop back to \textit{Define Hyper Space}) or b) by selecting more appropriate space filling or by increasing the surrogate model itself \cite{Krige1951} \cite{Cressie1990} (i.e. loop back to \textit{Design of virtual Experiments (DovE)}).This will allow an iterative optimization of both the model and the system itself. The surrogate model itself may be characterized by attributes such as validation area, p value \cite{Fisher1934} or other variables that may be interpreted as target indicators. When simultaneously interpreting the surrogate model parameter vector $\alpha$ as design variable the surrogate model may be optimized formally the same way as the system itself. The procedure is self-recursive.

\end{enumerate}

\begin{figure}[htbp]
\centerline{\includegraphics[width=8cm]{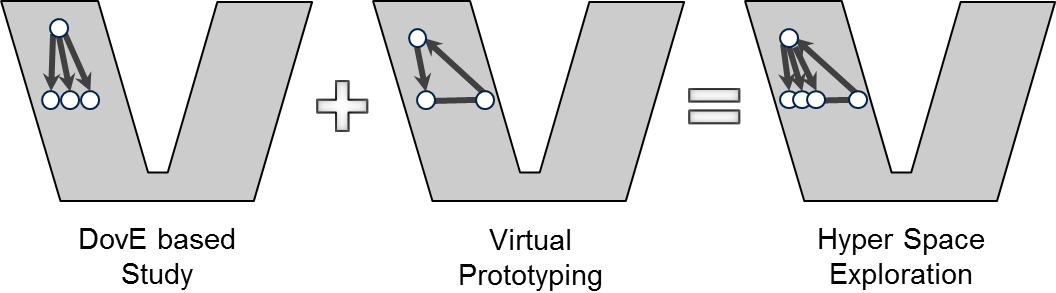}}
\caption{Schematic integrating the generic HSE work flow into the V-Model.}
\label{fig_HSE_V-Model_integration}
\end{figure}

The generic HSE work flow fits well into the V-Model. It may be considered as a combination of V-Model's macro-procedural elements (DovE based) study and (simulation based virtual) prototyping as indicated in Fig.~\ref{fig_HSE_V-Model_integration}.

\subsection{Generic HSE Tool Chain}
\begin{figure}[htbp]
\centerline{\includegraphics[width=7cm]{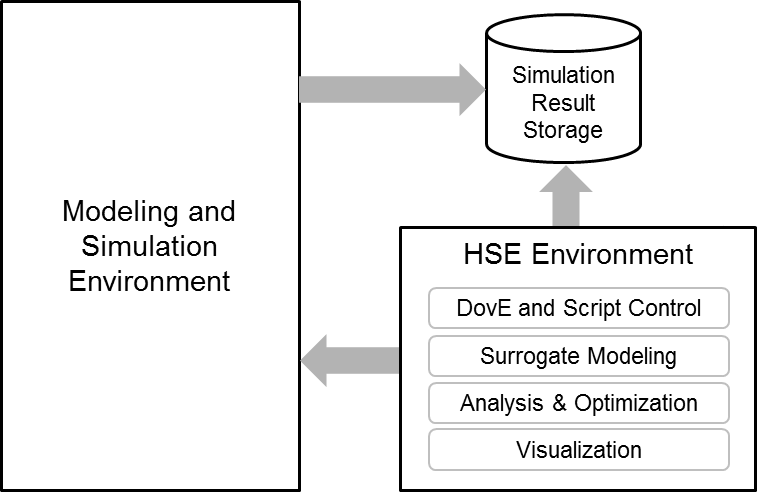}}
\caption{Generic HSE tool chain indicating its key components.}
\label{fig_generic_HSE_tool_chain}
\end{figure}

Executing the HSE work flow requests an existing tool chain as described in Fig. \ref{fig_generic_HSE_tool_chain} containing generic key components:
\begin{itemize}
\item The \textit{Modeling and Simulations Environment} is able to run domain or cross-domain specific individual simulations.
\item Simulation results are transferred to the \textit{Simulation Result Storage} giving access to further processing.
\item The \textit{HSE Environment} allows running HSE specific tasks of a) Hyper Space definition, DovE space filling and script control b) building a surrogate model c) analyzing and optimizing surrogate models and system designs and d) visualizing results for adequate analysis and decision making.
\end{itemize}

\section{Two exemplary HSE automotive applications}
\begin{figure}[htbp]
\centerline{\includegraphics[width=8cm]{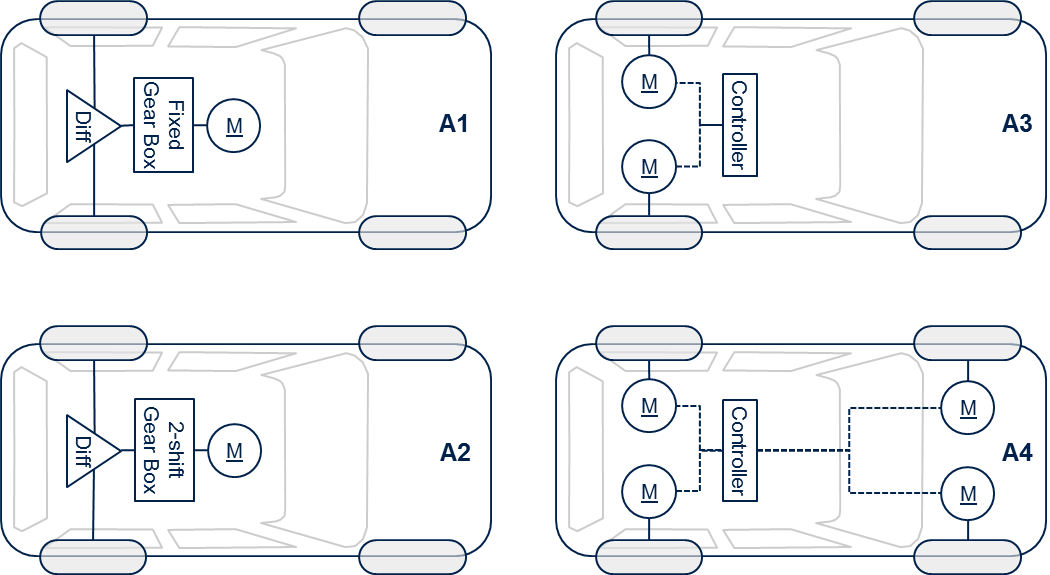}}
\caption{HSE evaluated topological and control alternatives for automotive design. Alternative A1 differs from A2 in terms of a fixed instead of a 2-shift gear box. Alternative A3 differs from A4 in terms of a 4-wheel instead of a 2-wheel drive train and an adjusted control strategy.}
\label{fig_HSE_ATV_alternatives}
\end{figure}

The HSE methodology has progressively evolved over the past years from a virtual prototyping based study to today's multicriterial quantitative trade-off analysis capabilities for system design in complex environment \cite{Schneider2012} \cite{Palm2013a} \cite{Palm2013b} \cite{Holzmann2016}. HSE development was constantly accompanied by industrial applications. Two examples of automotive applications with selected results already published in \cite{Palm2013b}  \cite{Holzmann2016} may demonstrate universality and mightiness of the HSE approach in the following two subsections. Both refer to the development of fully electric vehicles (FEVs). Example 1 follows the question if FEVs may or may not benefit from shiftable gear boxes. Alternative topologies A1 and A2 as shown in Fig. \ref{fig_HSE_ATV_alternatives} with varying component layouts, therefore, had to be compared.  

Example 2 follows the question of potential lateral vehicle stability benefit of 2-wheel versus 4-wheel drive trains in FEVs when making use of active yaw control \cite{Novellis2012} approaches. Alternative topologies A3 and A4 as shown in Fig. \ref{fig_HSE_ATV_alternatives} with varying controller layouts, therefore, had to be compared. 

\subsection{Do FEVs benefit from a shiftable gear box?}
\begin{figure}[htbp]
	\centerline{\includegraphics[width=8cm]{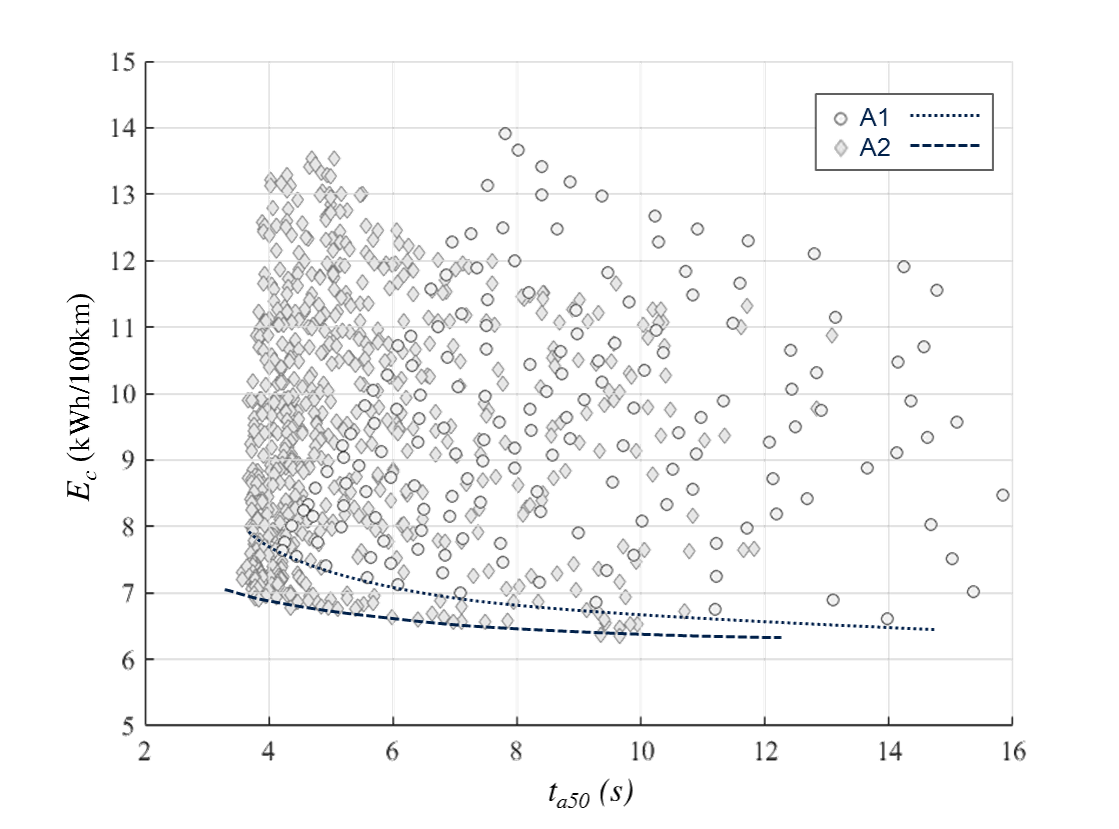}}
	\caption{HSE Trade-offs for A1 versus A2 design layout alternatives also indicating fronts of Pareto-optimal solutions for A1 and A2, respectively. Further modeling and simulation details may be found in \cite{Palm2013b}. }
	\label{fig_A1_A2_trade-off}
\end{figure}

Component parameter layouts (such as the electric engine's maximum torque) of FEV drive trains will significantly differ when considering a fixed gear box (Fig. \ref{fig_HSE_ATV_alternatives} alternative A1) instead of a shiftable gear box (Fig. \ref{fig_HSE_ATV_alternatives} alternative A2). While this fact usually leads to incommensurable design alternatives an HSE analysis allows direct comparison of their respective \textit{potentials}. Results of an HSE analysis A1 versus A2 are shown in Fig. \ref{fig_A1_A2_trade-off}. Each symbol represents a potential layout alternative. The two differing FEV topology approaches A1 and A2 are compared with respect to two relevant target indicators: An acceleration time $t_{a50}$ from zero to $50km/h$ and an energy consumption $E_{c}$ within the use-case of a NEDC \cite{NEDC} driving cycle normalized to $100km$ range. 

A dotted and a dashed line in Fig. \ref{fig_A1_A2_trade-off} represent the fronts of Pareto-optimal solutions for both alternatives A1 and A2, respectively. A 4th grade polynomial has been used as surrogate model base. Most layout alternatives represented in Fig. \ref{fig_A1_A2_trade-off} may be improved with respect to both target indicators. Comparing the two \textit{Pareto fronts}, however, indicates the fundamental difference between potentials of both architectural approaches.

\subsection{Are 2-wheel drives more stable than 4-wheel drives?}
\begin{figure}[htbp]
\centerline{\includegraphics[width=8cm]{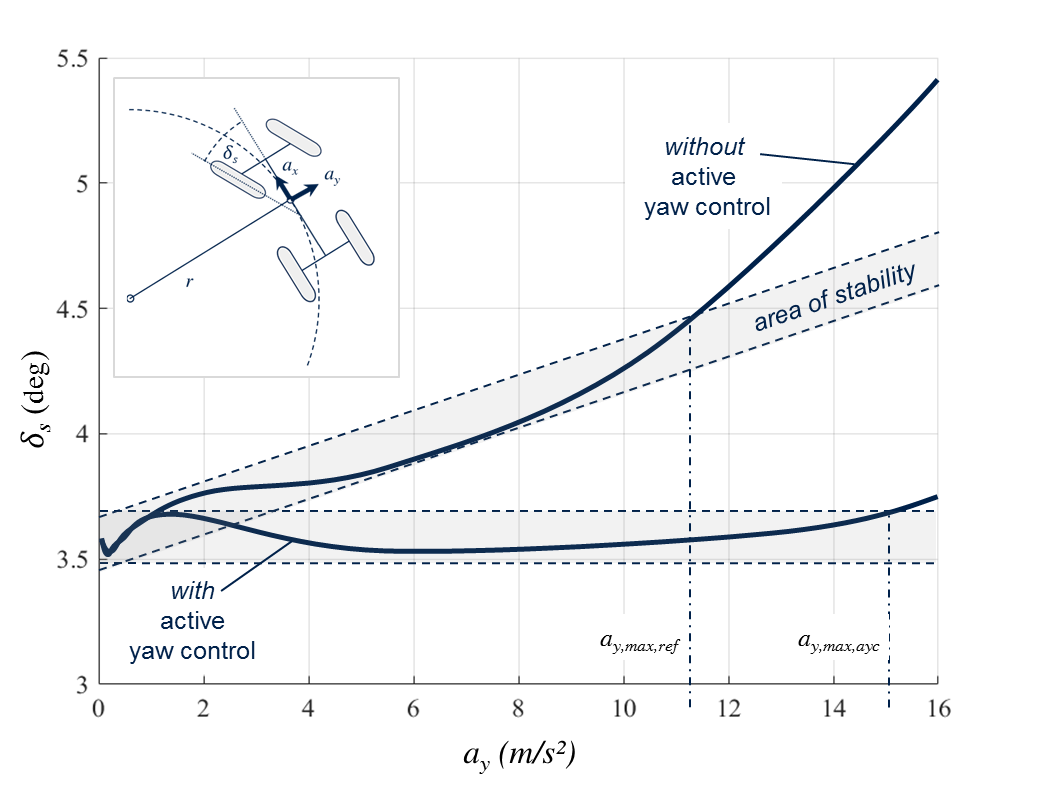}}
\caption{Principle scenario for quantifying a  vehicle's lateral stability.}
\label{fig_lateral_stability_measurement_scenario}
\end{figure}

Amongst many other topics automotive engineering is dealing with the question of how lateral vehicle stability may be achieved. In the inset of Fig. \ref{fig_lateral_stability_measurement_scenario} of a use case according to \cite{Novellis2015} is illustrated allowing to quantify the lateral stability criterion: When driving with constant longitudinal acceleration $a_x$ along a circle line of constant radius $r$ lateral acceleration $a_y$ will continually increase. Drivers have to adjust the vehicle steering angle $\delta_s$ in order to stay on track. As long as the driving angle may be kept within an area of stability (according to a linear steering behavior as indicated in Fig. \ref{fig_lateral_stability_measurement_scenario}) of constant width drivers may perceive vehicle behavior to be stable. Leaving the area of stability marks the maximum lateral acceleration value of stability. 

Active yaw control approaches \cite{Novellis2012} allow to change the $\delta_s (a_y)$ behavior and, thereby, shift the maximum value $a_{y,max}$. A target indicator  $gain_{stab}$  referring to the stability gain between the lateral acceleration maximum of stability with active yaw control ($a_{y,max,ayc}$) versus without active yaw control ($a_{y,max,ref}$)  may be defined as:

\begin{equation}
gain_{stab} := \frac{a_{y,max,ayc}}{a_{y,max,ref}} -1
\label{eq:a_ymax_definition}
\end{equation}

\begin{figure}[htbp]
\centerline{\includegraphics[width=8cm]{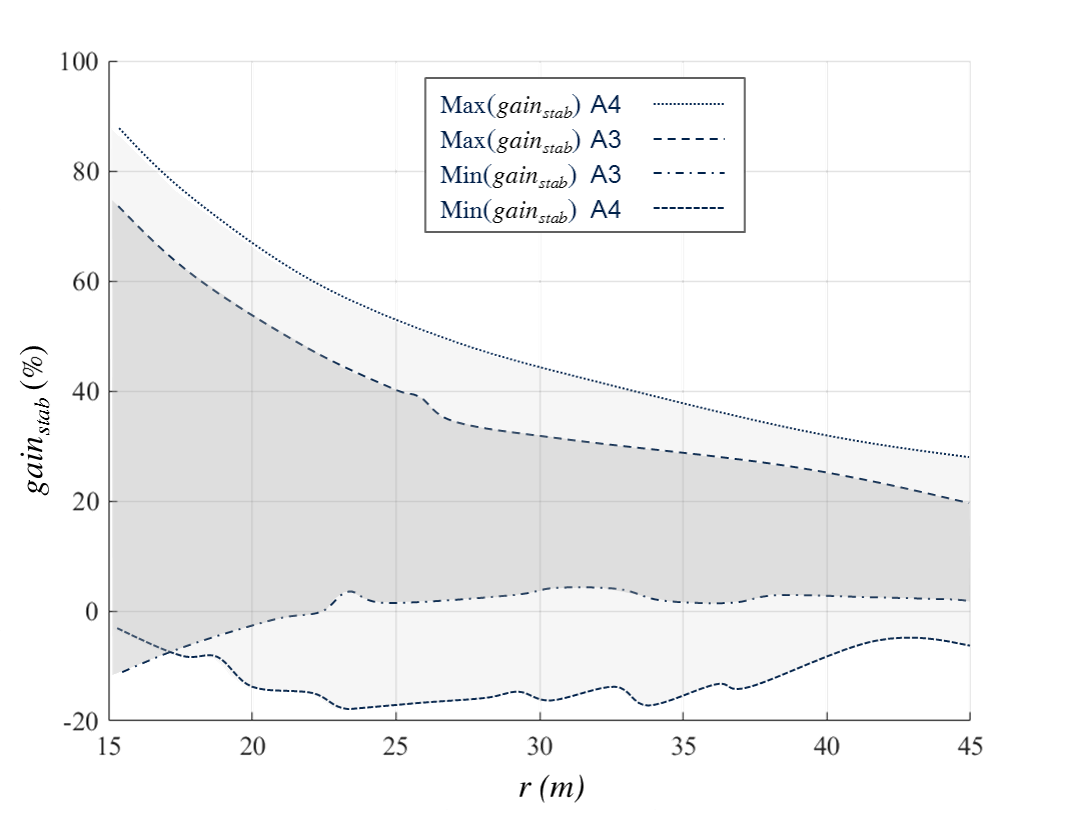}}
\caption{Stability gain analysis for of an active yaw control when used within an active 4-wheel (light gray area) vs. 2-wheel (dark gray area) drive train. Further modeling and simulation details may be found in \cite{Holzmann2016} .}
\label{fig_stability_gain_analysis_ayc}
\end{figure}

Designing an active yaw controller may become a complex task when extending the \textit{Design Space} (spanned by all controller variables) by the \textit{Use Case Space} (spanned in our example by the two use case variables $a_x$ and $r$). HSE allows a systematic approach for system optimization even within this environment of multiple scenarios. Fig. \ref{fig_stability_gain_analysis_ayc} shows the quantified dependency of the vehicle stability (expressed by target indicator $gain_{stab}$) as a function of road curvature (expressed by use case variable $r$) for \textit{all} considered controller layouts and use cases of longitudinal acceleration $a_x$. 

Fig. \ref{fig_lateral_stability_measurement_scenario} represents a potential analysis (best case vs. worst case ranges) of vehicle stability in all kinds of relevant design alternatives and use-case scenarios as defined within the \textit{Hyper Space}. In the chosen example, we can learn: A 4-wheel drive train at its Pareto-optimal active yaw controller layout may add an approximate 15\% lateral stability improvement when compared to a 2-wheel drive train also at Pareto-optimal active yaw controller layout. However, when the active yaw control is not well parameterized or suited for a specific use-case, 4-wheel drive trains may get even detrimental as compared to nonactive control. Loss of vehicle stability by a badly parameterized 2-wheel drive train active yaw controller, in contrast to that, is significantly lower.

\section{HSE Outlook}
The proposed HSE methodology is suited for system design in any complex environment. Architecting of systems may significantly benefit from HSE's multicriterial quantitative trade-off analysis and statistical learning. There are numerous systems in all potential application fields worthwhile to be considered. To name just two of them: a) Rebuilding energy supply systems for achieving sustainability bears a huge number of layout alternatives on any hierarchic level. Cost of implementation and time pressure are equally high as existing uncertainty to identify best solutions. b) Automated driving applications make massive use of artificial intelligence algorithms. So far a direct comparison of the \textit{potentials} of system architectures with their multitude of design alternatives (treating algorithm layouts the same way as component parameter or topological layouts) and use case variations is missing. HSE is capable to open a door for this task.  \newline
Besides additional applications there are still a few open questions to be answered for allowing comprehensive HSE application in any complex system environment. To name just one: While existing space filling algorithms bear no issue of dimensionality computational performance may still be a limiting factor when entering \textit{Hyper Spaces} with more than a hundred variables. HSE extending methods such as hierarchic search algorithms or model reduction approaches have to be investigated. \newline
HSE already proved its mightiness in architecting of complex systems. For becoming state-of-the-art fellow campaigners are highly welcome.

\section*{Acknowledgment}

The author of this review like paper thanks all contributors over the past years who helped to evolve Hyper Space Exploration from an idea to a respected application proven methodology.



\begin{thebibliography}{00}
\bibitem{Snow2007} David J. Snowden , Mary E. Boone: ``A Leader's Framework for Decision Making'', Harvard Business Review, 11/2007.
\bibitem{Cyn4Eng2017} Jan Vollmar, Michael Gepp, Herbert Palm, Ambra Cal\`{a}: ``Engineering framework for the future - Cynefin for Engineers'', IEEE International Symposium on Systems Engineering, Wien, 2017.
\bibitem{Boehm1979}  Boehm, Barry W.: ``Guidelines for Verifying and Validating Software Requirements and Design Speciﬁcations'', Euro IFIP 79, pp. 711-719, 1979
\bibitem{Dem1986} William E. Deming: ``Out of the crisis'', Cambridge, MA: Massachusetts Institute of Technology, Center for Advanced Engineering Study. p. 88. ISBN 0911379010. OCLC 13126265, 1986.
\bibitem{Fors1998}  Kevin Forsberg and Harold Mooz: ``System Engineering for Faster, Cheaper, Better'', Center for Systems Management, Reprinted by SF Bay Area Chapter of INCOSE (International Council of Systems Engineering), 1998.
\bibitem{Schaaf1997}  James Schaaf and Faye Lynn Thompson: ``Systems Concept Development with Virtual Prototyping'', Proceedings of the 
1997 Winter Simulation Conference, pp. 941-947.
\bibitem{Daenzer2002} Walter Daenzer and Fritz Huber (editors): ``Systems Engineering: Methodik und Praxis'', Industrielle Organisation, Zurich, 11th edition, 2002.
\bibitem{Ross2005} Adam Ross and Daniel Hastings: ``The Tradespace Exploration Paradigm'', INCOSE international Symposium, Vol 15, No. 1(1), 2005.
\bibitem{Zwicky1969} Fritz Zwicky ``Discovery, Invention, Research - Through the Morphological Approach'', The Macmillan Company, Toronto, 1969.
\bibitem{Dobbs2015}  Richard Dobbs, James Manyika and Jonathan Woetzel: ``No Ordinary Disruption: The Four Global Forces Breaking All the Trends'', Public Affairs, 2015.
\bibitem{SE2015} David Walden et al.:``INCOSE Systems Engineering Handbook: A Guide for System Life Cycle Processes and Activities'', Wiley, Hoboken, NJ, 4. ed. Auflage, ISBN 9781118999400, 2015.
\bibitem{NASA2007} National Aeronautics and Space Administration: ``NASA Systems Engineering Handbook'', Vol. 2007-6105, NASA / SP., Washington D.C, 1st edition, ISBN 9780160797477, 2007.
\bibitem{Kang2010} Eunsuk Kang, Ethan Jackson and Wolfram Schulte: ``An approach for effective design space exploration'', FOCS'10 Proceedings of the 16th Monterey conference on Foundations of computer software: modeling, development, and verification of adaptive systems, Springer-Verlag Berlin, Heidelberg, ISBN 978-3-642-21291-8, 2010.
\bibitem{Neufville2010} Richard Neufville: ``Real Options: Dealing With Uncertainty in Systems Planning and Design'', Journal of Integrated Assessment, Volume 4, 2003
\bibitem{Fisher1934} Sir Ronald Aylmer Fisher: ``The Design of Experiments'', published by Oliver and Boyd, 1935.
\bibitem{Box2005} George Box, Stuart Hunter and William Hunter: ``Statistics for Experimenters'' (2nd ed.), John Wiley \& Sons, 2005.
\bibitem{Forr2008} Alexander Forrester, Andr\`{a}s S\^{o}bester und Andy Keane: ``Engineering design via surrogate modelling: A practical guide'', J. Wiley, Chichester, West Sussex, England, Hoboken, NJ, ISBN 978-0-470-06068-1, 2008.
\bibitem{Krige1951} Danie Krige: ``A Statistical Approach to Some Basic Mine Valuation Problems on the Witwatersrand''. Journal of the Chemical, Metallurgical and Mining Society of South Africa, Vol. 52:119–139, 1951.
\bibitem{Cressie1990} Noel Cressiel: ``The origins of kriging'', Mathematical Geology, 22(3):239–252, ISSN 0882-8121, 1990.
\bibitem{Schneider2012} Stefan-Alexander Schneider, Bernhard Schick, Herbert Palm: ``Virtualization, Integration and Simulation in the Context of Vehicle Systems Engineering'', Embedded World 2012, N\"{u}rnberg, Germany; 02/2012.
\bibitem{Palm2013a} Herbert Palm, J\"{o}rg Holzmann, Robert Klein, Dieter Gerling: ``A Novel Approach on Virtual Systems Prototyping Based on a Validated, Hierarchical, Modular Library'', Embedded World 2013, N\"{u}rnberg; 02/2013.
\bibitem{Palm2013b} Herbert Palm, J\"{o}rg Holzmann, Stefan-Alexander Schneider, Hans-Michael K\"{o}geler: ``The Future of Car Design - Systems Engineering based Optimisation'' Springer, ATZ 06/2013, 115(6):512-517, 2013.
\bibitem{Holzmann2016} J\"{o}rg Holzmann, Herbert Palm and Dieter Gerling: ``Virtual Prototyping basierte Trade-off Analysen", „Tag des Systems Engineering“, S.99-108, ISBN: 978-3-446-45126-1, Hrsg.: Schulze, S.O., Muggeo C., Hanser Verlag, 2016.
\bibitem{Novellis2012} Leonardo De Novellis et al.: ``Torque Vectoring for Electric Vehicles with Individually Controlled Motors: State-of-the-Art and Future Developments", EVS26 International Battery, Hybrid and Fuel Cell Electric Vehicle Symposium, Los Angeles, California, May 6-9, 2012.
\bibitem{NEDC} E/ECE/324/Rev.2/Add.100/Rev.3, 2013.
\bibitem{Novellis2015} Leonardo De Novellis, Aldo Sorniotti, PatrickGruber: ``Driving modes for designing the cornering response of fully electric vehicles with multiple motors", Mechanical Systems and Signal Processing 64-65, pp. 1-15, 2015.

\end{thebibliography}
\end{document}